\documentclass[preprint,showpacs,floatfix,superscriptaddress,nofootinbib]{revtex4}
\usepackage{graphicx}
\usepackage{epsfig}
\usepackage{bm}
\usepackage{amsfonts}
\usepackage[T1]{fontenc}
\usepackage{amssymb}
\usepackage{color}
\usepackage{float}
\usepackage{amsmath}
\usepackage{urwchancal}
\usepackage{dcolumn}
\usepackage{cancel}
\usepackage[colorlinks]{hyperref}
\hypersetup{
 breaklinks=true,
 pdfstartview={FitH}, % fits the width of the page to the window
 colorlinks=true, % false: boxed links; true: colored links
 linkcolor=blue, % color of internal links
 citecolor=red, % color of links to bibliography
 filecolor=magenta, % color of file links
 urlcolor=blue, % color of external links
 anchorcolor=green, % Color for anchor text
 linktocpage=true
}
\usepackage{microtype}
\begin{document}

\title{Thermodynamics of $2+1$ dimensional Coulomb-Like Black Holes under Nonlinear Electrodynamics with a traceless energy-momentum tensor}

\author{Mauricio Cataldo}
\email{mcataldo@ubiobio.cl} \affiliation{Departamento de F\'{\i}sica, Grupo Cosmolog\'{\i}a y Part\'{\i}culas Elementales, Universidad del B\'{\i}o-B\'{\i}o, Casilla 5-C, Concepci\'on, Chile..}
\author{P. A. Gonz\'{a}lez}
\email{pablo.gonzalez@udp.cl} \affiliation{Facultad de
Ingenier\'{i}a y Ciencias, Universidad Diego Portales, Avenida Ej\'{e}rcito
Libertador 441, Casilla 298-V, Santiago, Chile.}
\author{Joel Saavedra}
\email{joel.saavedra@ucv.cl} \affiliation{Instituto de
F\'{i}sica, Pontificia Universidad Cat\'olica de Valpara\'{i}so,
Casilla 4950, Valpara\'{i}so, Chile.}
\author{Yerko V\'{a}squez}
\email{yvasquez@userena.cl}\affiliation{Departamento de F\'{\i}sica, Facultad de Ciencias, Universidad de La Serena,\\ 
Avenida Cisternas 1200, La Serena, Chile.}
\author{Bin Wang}
\email{wang_b@sjtu.edu.cn}\affiliation{School of Aeronautics and Astronautics, Shanghai Jiao Tong University, \\ Shanghai 200240, China.}
\affiliation{Center for Gravitation and Cosmology, College of Physical Science
and Technology, Yangzhou University,\\ Yangzhou 225009, China.}

\date{\today}

\begin{abstract}

 In this work we study the thermodynamics of a (2+1)-dimensional static black hole under a nonlinear electric field. In addition to standard approaches, we investigate black hole thermodynamic geometry. We compute Weinhold and Ruppeiner metrics and compare the thermodynamic geometries with the standard description for black hole thermodynamics. We further consider the cosmological constant as an additional extensive thermodynamic variable. For thermodynamic equilibrium in three dimensional space, we compute heat engine efficiency and show that it may be constructed with this black hole.
\end{abstract}
\maketitle

\section{introduction}
Three-dimensional models of gravity have been of great interest due to their simplicity over four-dimensional and higher-dimensional models of gravity, and since some of the properties shared by their higher dimensional analogs can be more efficiently investigated.
In this sense, the well-known Bañados–Teitelboim–Zanelli (BTZ) black hole \cite{Banados:1992wn} - a solution to the Einstein equations in three dimensions with a negative cosmological constant - shares several features of Kerr black holes \cite{Carlip:1995qv}. Furthermore, topologically massive gravity (TMG) - constructed by adding a gravitational Chern-Simons term to the action of three-dimensional GR \cite{Deser:1981wh} and which contains a propagating degree of freedom in the form of a massive graviton \cite{Deser:1981wh, Deser:1982vy} - also admits the BTZ (and other) black holes as exact solutions \cite{Moussa:2003fc, Garbarz:2008qn, Vasquez:2009mk}.
 
Next, Bergshoeff, Hohm and Townsend presented the standard Einstein-Hilbert term with a specific combination of the scalar curvature square term and the Ricci tensor square one, known as BHT massive gravity \cite{Bergshoeff:2009hq, Bergshoeff:2009aq, Bergshoeff:2009tb,Bergshoeff:2009fj,Andringa:2009yc,Bergshoeff:2010mf}. BHT massive gravity admits interesting solutions \cite{Clement:2009gq, AyonBeato:2009yq, Clement:2009ka,AyonBeato:2009nh,Oliva:2009ip},
for further aspects see \cite{Kim:2009jm,Oda:2009ys, Liu:2009pha, Nakasone:2009vt, Deser:2009hb}. Although one might today consider Lorentz invariance neither fundamental nor exact, by introducing a preferred foliation and terms that contain higher-order spatial derivatives, significantly improved UV behavior can be achieved through what is known as Hořava gravity \cite{Horava:2009uw}. This theory admits a Lorentz-violating version of the BTZ black hole, as well as black holes with positive and vanishing cosmological constant \cite{Sotiriou:2011dr,Sotiriou:2014gna}.
Also, three-dimensional theories of gravity allow GR coupled to electromagnetic fields - be they Maxwell electrodynamics or nonlinear, such as Born-Infeld electrodynamics \cite{Born}. Here, Born-Infeld gravity has been a growing field of research, widely applied to black holes. Finally, there are many exact solutions of charged black holes under different frameworks considering general relativity or modified gravity, see \cite{Dey:2004yt,Cai:2004eh,Zou:2013owa,Boillat:1970gw,Fernando:2003tz,Jing:2010zp,deOliveira:1994in,Gullu:2010pc,Hendi:2014mba,Hendi:2016pvx,Hendi:2017mgb,Hendi:2017oka,Hendi:2020yah} and references therein. One remarkable solution corresponds to regular charged black holes, found by Ayon-Beato and Garcia
in Ref. \cite{Eloy}.

Given the above precedents, this work considers a three-dimensional static black hole solution that arises from nonlinear electrodynamics, and that satisfies weak
energy conditions. The electric field $E(r)$ is given by $E(r) = q/r^2$, and thus takes Coulomb's form for a point charge in Minkowski spacetime. The solution describes charged (anti)–de Sitter spacetimes \cite{Cataldo:2000we}. We explore the general formalism for such black holes, and disclose the corresponding thermodynamic properties through both the standard and geometrothermodynamics approaches. 

Nonlinear electrodynamics provides interesting solutions to the self-energy of charged point-like particles in Maxwell's theory \cite{Born}, and is of great interest in the context of low energy string theory \cite{Leigh:1989jq,Fradkin}. Moreover, nonlinear field theories are of interest to different branches of mathematical physics because most physical systems are inherently nonlinear in nature. Extending black holes from Maxwell fields to nonlinear electrodynamics can help to better understand the nature of different complex systems. Also, considering that three dimensional black holes help us find a profound insight in the quantum view of gravity, the generalization of the charged BTZ black hole to non-linear electrodynamics can help to obtain deeper insights into more information in quantum gravity. It was even argued that the effect of higher order corrected Maxwell field may compare with the effect brought by the correction in gravity.
On the other hand, it is known that when the 2+1 gravity is coupled to the Maxwell electromagnetic field, the solution is the usual charged BTZ black hole, which is not a spacetime of constant curvature and there is a logarithmic function in the metric expression, which makes the analytic investigation difficult and leads the spacetime not thoroughly investigated. Now, introducing the nonlinear electrodynamics as the source of the Einstein equation, a (2+1)-static black hole solution with a nonlinear electric field was obtained \cite{Cataldo:2000we}, that is a spacetime of constant curvature, which has the Coulomb form of a point charge in the Minkowski spacetime, instead of a logarithmic electric potential, keeping the mathematical simplicity as in the (3+1) gravity. Additionally, considering that three-dimensional black holes help us find a profound insight in the quantum view of gravity, the generalization of the charged BTZ black hole to non-linear electrodynamics can help to obtain deeper insights into more information in quantum gravity. 
As thermodynamic objects, black holes and their properties have been the subject of growing research since the seminal laws of black hole mechanics were published \cite{laws}. From a modern point of view, the study of the thermodynamic properties of AdS black holes provides important insight into AdS/CFT conjectures and, more recently, de Sitter/conformal field theory correspondence (dS/CFT) \cite{strominger}. In building upon standard descriptions, Weinhold introduced the first concepts of geometry into understanding thermodynamics, presenting a Riemannian metric as a function of the second derivatives
of the internal energy with respect to entropy and other extensive quantities of a thermodynamic system \cite{W}. Ruppeiner further explored geometrical concepts, defining another metric as the second derivative of entropy with respect to the internal energy and other extensive quantities of a thermodynamic system \cite{R}. This latter is conformally related to the Weinhold metric by the inverse temperature. The Ruppeiner geometry has its physical meanings in the fluctuation theory of equilibrium thermodynamics \cite{Ruppeiner:1995zz}, and was recently suggested as a possible means to disclose the microscopic structures of a black hole system \cite{Wei:2015iwa}. Thus it is interesting to compare the standard and geometrical methods to better understand black hole thermodynamics. To this end, we will explore in this paper the thermodynamic geometry of a 2+1 dimensional AdS black hole under a nonlinear electric field. 

The study of black hole thermodynamics has been generalized to the extended phase space, where the cosmological constant is identified with thermodynamic pressure and its
variations are included in the first law of black hole thermodynamics (for a review, please refer to \cite{Kubiznak:2016qmn}). In the extended phase space - with cosmological constant and volume as thermodynamic variables - it was interestingly found that
the system admits a more direct and precise coincidence between the first order small-large black hole phase transition and the liquid-gas change of phase occurring in fluids \cite{Kubiznak:2012wp}. Considering the extended phase space, and hence treating the cosmological constant as a dynamical quantity, is a very interesting theoretical idea in disclosing possible phase transitions in AdS black holes \cite{ Gunasekaran:2012dq}. More discussions in this direction can be found in existing references. The thermodynamics of D-dimensional Born-Infeld AdS black holes in the extended phase space was examined in \cite{Zou:2013owa}. Here we generalize the extended phase space thermodynamics discussion to 2+1 dimensional AdS black holes under a nonlinear electric field. We examine whether critical behavior in the extended phase space thermodynamics displays special properties.

The paper is organized as follows. In Sec. \ref{background} we give a brief review of the literature. In Sec. \ref{termo}, we study thermodynamics of the spacetime following standard and geometrothermodynamics approaches. In Sec. \ref{ETD}, we generalize our discussion to extended space thermodynamics. Finally, we conclude and discuss the results obtained in Sec. \ref{conclusion}.

%%%%%%%%%%%%%%%%%%%%%%%%%%%%%%%%%%%%%%%%%%%%%%%%%%%%%%%%%%%%%%%%%%

\section{General Formalism for 2+1 Dimensional Gravity under Nonlinear Electrodynamics}
\label{background}
We consider the black hole solution to arise from nonlinear electrodynamics 
 \begin{eqnarray} \label{action}
 S=\int d^{3}x\sqrt{-g}\left(\frac{1}{16 \pi}\left(R-2 \Lambda\right)+L(F)\right)~,
 \end{eqnarray}
where L(F) represents the electromagnetic Lagrangian for nonlinear electrodynamics, the electromagnetic tensor is written in the usual form from the vector potential 
$A_{\mu}$, and the electromagnetic tensor as $F_{\mu \nu}=\partial_\mu A_\nu- \partial_\nu A_\mu$. The variation in respect to metric ($\delta g_{\mu \nu}$) and vector potential ($\delta A_\mu$) gives the equation of motion

 \begin{eqnarray}
 G_{\mu\nu}=-\Lambda g_{\mu\nu}+8\pi T_{\mu\nu}\,,
 \label{field1}
 \end{eqnarray}
where the energy-momentum tensor $T_{\mu\nu}$ for the
electromagnetic field is given by
 \begin{eqnarray}
 T_{\mu\nu} = g_{\mu\nu}L(F)-F_{\mu \rho}F_\nu^\rho L_{,F}\label{energymomentum}~.
 \end{eqnarray}
 The electromagnetic equation is given by
 \begin{eqnarray}
 \nabla_\mu \left( F^{\mu \nu} L_{,F}\right)=0\,,\label{maxwel1}
 \end{eqnarray}
 The solution to these equations for a vanishing trace energy-momentum tensor is given in \cite{Cataldo:2000we}.
 It is well known that the electromagnetic energy-momentum tensor in $3+1$ Maxwell electrodynamics is trace free, given by $T=T_{\mu \nu}g^{\mu \nu}$, with standard Coulomb solution. In contrast, for $2+1$ dimensions, the trace of the electromagnetic energy-momentum tensor is not vanished. Under Maxwell theory, a $2+1$ always has trace, and the electric field for a circularly symmetric static metric coupled to a Maxwell field is proportional to the inverse of $r$, i.e., $E_r\propto 1/r$. Hence the vector potential $A_0$ is logarithmic, i.e., $A \propto ln r$ and consequently blows up at $r=0$. In order to find physical quantities like mass, electric charge, etc., we need to introduce a renormalization scheme.
 
 This article focuses on electromagnetic theories where the main condition is having a traceless energy-momentum tensor. Of course, this condition restricts the class of nonlinear electrodynamics under consideration. Moreover, if we demand this condition from the electromagnetic domain of equation (\ref{action}), Ref. \cite{Cataldo:2000we} showed that it may only be fulfilled under a Lagrangian proportional to $F^{3/4}$. Now, in this case - a circularly symmetric static metric - the resulting solution for the electric field is proportional to the inverse of $r^2$, surprisingly alike the Coulomb law for a point charge in $3+1$ dimensions. Furthermore, the energy-momentum tensor satisfies the weak energy condition. Consider the following metric ansatz
\begin{eqnarray}
 ds^{2}=-f(r)dt^{2}+f^{-1}(r)dr^{2}+r^2 d \phi^2\,. \label{metricBH}
 \end{eqnarray}
Now, to summarize the main results of Maxwell equation (\ref{maxwel1}) under the condition of vanished trace, we have
\begin{equation}
 T=T_{\mu \nu}g^{\mu \nu}=3L(F)-4FL,_F\,,
\end{equation}
which yields 
\begin{equation}
 L(F)=C|F|^{3/4}\,,
\end{equation}
 where $C$ is an integration constant. Because the magnetic field is vanished as a consequence of Einstein's equation, we get
 \begin{equation}\label{F}
 L(F)=C E^{3/2}\,,
\end{equation}
from Maxwell equation
 \begin{equation}
 \frac{d}{dr}\left ( r E L,_F\right)=0\,,
\end{equation}
which, integrated, gives 
 \begin{equation}
 E(r) L_{,F}=-\frac{q}{4 \pi r}\,,
\end{equation}
where $q$ is an integration constant. From (\ref{F}) it follows that
 \begin{equation}
 E(r)=\frac{q^2}{6 \pi C}^2\frac{1}{r^2}\,,
\end{equation}
and finally, setting $C=\sqrt{|q|}/6 \pi$, the electric field becomes
\begin{equation}
 E(r)=\frac{q}{r^2}\,.
\end{equation}

\begin{figure}[!h]
	\begin{center}
		\includegraphics[width=80mm]{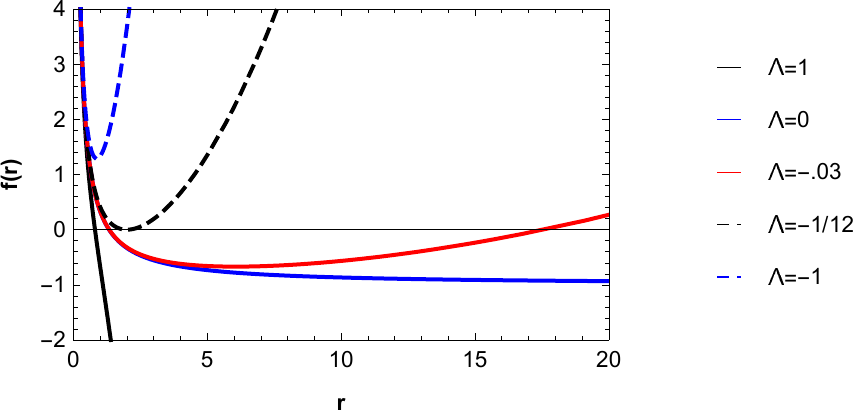}
	\end{center}
	\caption{The behavior of metric function $f(r)$, with $M=1$, $q=1$, for different values of cosmological constant. %Here we can see the different behavior depending of the sign of $\Lambda$. $\Lambda>0$ and $\Lambda=0$ we have a cosmological horizon, $\Lambda>0$ there are different ranges where we have two horizon black hole, single or naked singularities.
	Note that when $\Lambda=-1$, there is a naked singularity.
	}
	\label{lapsus1}
\end{figure}

Now, under the traceless condition, the components of Einstein equations $R_{tt}=(-f^2 R_{rr})$ and $R_{\omega \omega}$ can be written as
\begin{eqnarray}
f_{,rr}+\frac{f_{,r}}{r}&=&-2 \Lambda +\frac{2q^2}{3r^2}\,,\label{Rtt}\\
f_{,r}&=&-2 \Lambda-\frac{4q^2}{3r^2}\,. \label{Rww}
\end{eqnarray}
It is easy to show Eq. (\ref{Rtt}) by virtue of the Maxwell equations. Therefore the only remaining component of Einstein equations (\ref{Rww}) can be directly integrated, with the lapse function given by 
 \begin{equation}
 f(r)=-M-\Lambda r^2+\frac{4q^2}{3r}\,, \label{f(r)}
 \end{equation}
 where $M$ is a constant related to the physical mass and $q$ is a constant related to physical charge. We will return to this point later to discuss the physical meaning of these constants.
Here, for $\Lambda > 0$ - i.e., asymptotically de-Sitter spacetime - the solution shows a cosmological horizon; under a vanishing cosmological constant, it has an asymptotically flat solution coupled with a Coulomb-like field that also shows a cosmological horizon; and under negative cosmological constant ($\Lambda <0$), we have a genuine black hole solution where its horizon corresponds to the solution of $f(r)=-M-\Lambda r_{h}^2+\frac{4q^2}{3r_{h}}=0$, given by
\begin{equation}
r_{h_{1}}=\frac{h}{3 \Lambda}-\frac{M}{h}~,\,\,\label{rh}
\end{equation}
\begin{equation}
r_{h_{2}}=-\frac{h}{6\Lambda}+\frac{M}{2h}+i\frac{\sqrt{3}}{2}\left(\frac{h}{3 \Lambda}+\frac{M}{h}\right)~,\,\,
\end{equation}
\begin{equation}
r_{h_{3}}=-\frac{h}{6\Lambda}+\frac{M}{2h}-i\frac{\sqrt{3}}{2}\left(\frac{h}{3 \Lambda}+\frac{M}{h}\right),
\end{equation}
where $h$ is 
\begin{equation}
h=\left(\left(18q^2+3\sqrt{3\left(\frac{M^3}{\Lambda}+12q^4\right)}\right)\Lambda^2\right)^{\frac{1}{3}}~.
\end{equation}
Figure \ref{lapsus1} shows the behavior of the lapse function at different cosmological constant values, where the horizon depends on the sign of $\Lambda$. Thus, if $\Lambda>0$ or $\Lambda=0$, there is a cosmological horizon; and if $\Lambda<0$, there are different ranges for two or single horizon black holes, or naked singularities.
Let us now consider the AdS case. Exploring black hole descriptions in (\ref{f(r)}), we will show how the sign of the radical is crucial to different solutions of
\begin{equation}
\alpha:=\frac{M^3}{\Lambda}+12q^4~,\,\,
\end{equation}
For $M>0$, the solution is essentially related to comparisons of the cosmological constant, black hole mass, and electric charge. So:
\begin{itemize}
\item Case $\alpha < 0$ or $ 0>\Lambda >- \frac{M^3}{12q^4}$, we get real solutions for the event horizon, and the solution represents a black hole with inner and outer horizons. This behavior is shown in Fig. \ref{lapsus2}.
\begin{figure}[!h]
	\begin{center}
		\includegraphics[width=80mm]{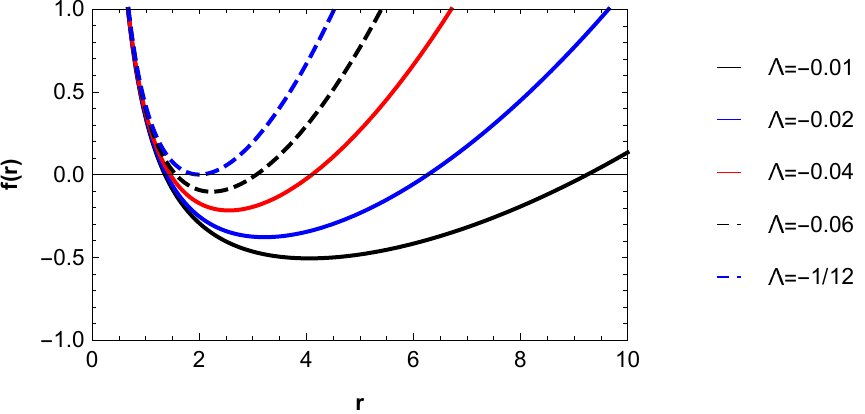}
	\end{center}
	\caption{The behavior of the metric function $f(r)$, with $M=1$, $q=1$, for different values of the cosmological constant $ 0>\Lambda >- \frac{M^3}{12q^4}$. Here, we observe a black hole with two horizons. Note the extremal configuration at $\Lambda=-\frac{1}{12}$.}
	\label{lapsus2}
\end{figure}
\newpage
\item Case $\alpha > 0$ or $ \Lambda <- \frac{M^3}{12q^4}$. There is one real and two complex solutions. Their behavior is shown in Fig. \ref{lapsus3}, generally naked singularities.
\begin{figure}[!h]
	\begin{center}
		\includegraphics[width=80mm]{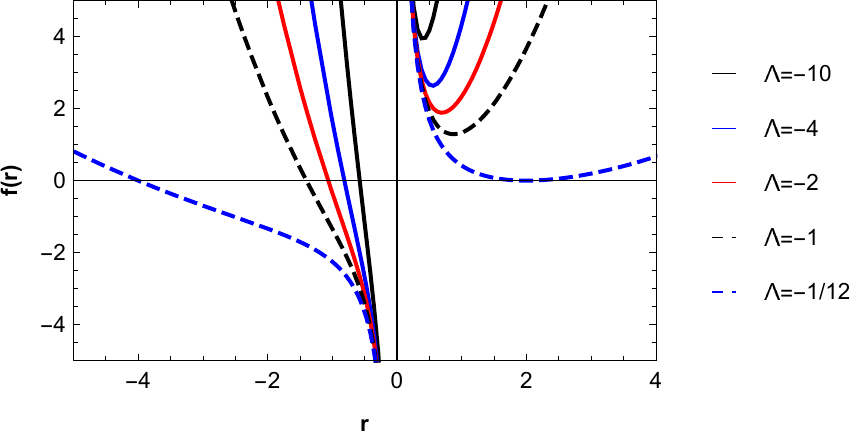}
	\end{center}
	\caption{The behavior of the metric function $f(r)$, with $M=1$, $q=1$, for different values of the cosmological constant when $ \Lambda <- \frac{M^3}{12q^4}$. The region is naked singularities: one real negative solution for the horizon, and two complex solutions. The plot includes extremal case $\Lambda=-\frac{1}{12}$. }
	\label{lapsus3}
\end{figure}
\item Case $\alpha =0 $ or $ \Lambda =- \frac{M^3}{12q^4}$, we get one real root solution, representing an extremal black hole. Figs. \ref{lapsus2} and \ref{lapsus3} show the extremal solutions in both cases.
\end{itemize}
The black hole solution is singular only at $r=0$. The first two invariant curvatures are
\begin{eqnarray}
R&=&6 \Lambda, \\
R_{\mu \nu}R^{\mu \nu}&=& 12 \Lambda^2+\frac{8q^4}{3r^6},
\end{eqnarray}
As mentioned above, there is a genuine singularity at the origin; note that these invariants are not singular at the horizon.

In finishing this section, we would like to do a comparison with charged BTZ black holes regarding the major differences of both solutions in $2+1$ dimensions with linear and nonlinear electrodynamic interactions. Briefly, the BTZ black hole is a solution to Einstein-Maxwell theory in AdS spacetime:
\begin{equation}
 f(r)= -2m +\frac{r^2}{l^2}-\frac{q^2}{2}ln\left(\frac{r}{l}\right), \label{BTZ}
\end{equation}
where $q$ and $m$ are the black hole charge and mass, respectively; $\Lambda = -\frac{1}{l^2}$ is the cosmological constant; and $l$ the AdS radius \cite{Banados:1992wn}. This case is very challenging to address: the asymptotic structure renders computation of the mass more problematic, and requires renormalization. Following the renormalization scale proposed in \cite{cadoni}, the solution reads as follows:
\begin{equation}
 f(r)= -2m_0+\frac{r^2}{l^2}-\frac{q^2}{2}ln\left(\frac{r}{r_0}\right), \label{BTZ2}
\end{equation}
where $m_0=m+\frac{q^2}{2}ln\left(\frac{r}{r_0}\right)$. \\

\section{A review of thermodynamic approaches}
\label{termo}

\subsection{The Euclidean action}
We may compute quasilocal energy and mass \cite{Cataldo:2000we} from (\ref{action}), considering $\mathcal{L}(F) = C |F| ^n$, $F= \frac{1}{4} F_{\mu \nu} F^{\mu \nu}$ and $n=3/4$, given by the Euclidean continuation of the metric 
\begin{equation} \label{solution}
ds^2= N(r)^2 f(r) d\tau^2+ \frac{dr^2}{f(r)} + r^2 d\phi^2 \,, \,\,\,\, A_t(r) = -q/r \,,
\end{equation}
where $\tau= i t$ is Euclidean time. The constant $C$ is set to the value
\begin{equation}
C= \frac{8 |q| ^{1/2}}{2^{1/4} 3 \pi}\,,
\end{equation}
given which the metric function takes the value
\begin{equation}
f(r)=- \Lambda r^2-M+\frac{4q^2}{3r} \,,
\end{equation}
and the conjugate momentum is given by
\begin{equation}
\pi^r = sgn(F) \sqrt{g}C n |F|^{n-1} F^{t r} \,.
\end{equation}
Therefore, the Euclidean action is given by
\begin{widetext}{\bf{
\begin{equation}
I_E = 2\pi \beta \int_{r_h}^{\infty} dr \left[ N\left( \frac{1}{2 \pi} (f'(r) + 2\Lambda r) - \frac{(1-2n)[ |\pi^r|]^{2n/(2n-1)}}{2n\left(\frac{Cnr}{2^{n-1}} \right)^{1/(2n-1)}} \right) - A_{t} \partial_r \pi^r \right]+ B \,.
\end{equation}}}
\end{widetext}
Now, varying with respect to $N$, $f$, $\pi^r$ and $A_{t}$ we obtain field equations
\begin{equation}
 \frac{1}{2 \pi} (f'(r) + 2\Lambda r) - \frac{(1-2n)[(\pi^r)^2]^{n/(2n-1)}}{2n\left(\frac{Cnr}{2^{n-1}} \right)^{1/(2n-1)}} = 0
 \,,
 \end{equation}
 \begin{equation}
 N'(r)=0\,,
 \end{equation}
 \begin{equation}
 A_t' + sgn (\pi^r) N \left( \frac{2^{n-1} \pi^r }{ Cn r} \right)^{1/(2n-1)} =0 \,, 
 \end{equation}
 \begin{equation}
 \partial_r \pi^r =0 \,,
 \end{equation}
Without loss of generality, we can set $N(r)=1$ by coordinate transformation, and for which $\pi^r$ is a constant. These equations are consistent with the field equations in (\ref{action}) and their solution in (\ref{solution}).
The boundary term of the Euclidean action is given by
\begin{equation}
\delta B = - 2 \pi \left[ \frac{1}{2 \pi} \delta f + A_t \delta \pi^r \right]_{r_h}^{\infty} \,.
\end{equation}
the variations of the field solutions at infinity are given by
\begin{equation}
\delta f |_{\infty} = - \delta M \,, \,\,\,\, A_t \delta \pi^r |_{\infty} =0 \,,
\end{equation}
and at the horizon by
\begin{equation}
\delta f |_{r_h} = -f' |_{r_h}\delta r_h = -\frac{4 \pi}{\beta} \delta r_h \,, \,\,\,\, A_t \delta \pi |_{r_h} = -\Phi \delta \pi |_{r_h}\,,
\end{equation}
where $\Phi= A_t(\infty)- A_t(r_h) = \frac{q}{ r_h}$ is the electric potential. Then, we obtain
\begin{equation}
I_E = \beta M - 4 \pi r_h + 2 \beta A_t(r_h) \pi^r \,,
\end{equation}
and the mass, entropy, and electric charge are respectively given by
\begin{eqnarray}
\mathcal{M} &=& \frac{\partial I_E}{\partial \beta} |_{\phi} - \frac{\phi}{\beta} \frac{\partial I_E}{\partial \phi} |_{\beta} = M\,, \\
\mathcal{S} &=& \beta \frac{\partial I_E}{\partial \beta} |_{\phi} -I_E = 4 \pi r_h\,, \\
\mathcal{Q} &=& -\frac{1}{\beta}\frac{\partial I_E}{\partial \phi} |_{\beta} = -2 \pi \pi^r\,,
\end{eqnarray}
where 
\begin{equation}
\pi^r= - sgn(q) 2^{1/4} C n |q|^{1/2}=- \frac{2 q }{\pi}\,.
\end{equation}
Finally, we obtain electric charge $\mathcal{Q}= 4 q$.
Thus, we can conclude that the analogous ADM mass is defined to be $M(\infty):=M$, and therefore we do not need an extra renormalization procedure. Essentially, the inclusion of the nonlinear term for the Maxwell Lagrangian interacting with gravity in $2+1$ dimensions acts as a regulator, and the problems arising from the logarithmic term disappears. 
Notably, the charged BTZ solution exhibits a similar behavior, i.e., black holes with two horizons, naked singularities, and the presence of extremal solutions depending on charge and mass values. In the following section, we study the thermodynamics properties of Coulomb-like black holes, and see the main differences with the thermodynamics description of BTZ black holes. Particularly, we explore the possibility of phase transition, reverse isoperimetric inequality, thermodynamics curvature, extended thermodynamics, and Coulomb-like black holes as heat engines.

\subsection{Thermodynamics properties under the standard approach}
\begin{figure}[!h]
\begin{center}
%\plotone{Figg1}
\includegraphics[width=0.45\textwidth]{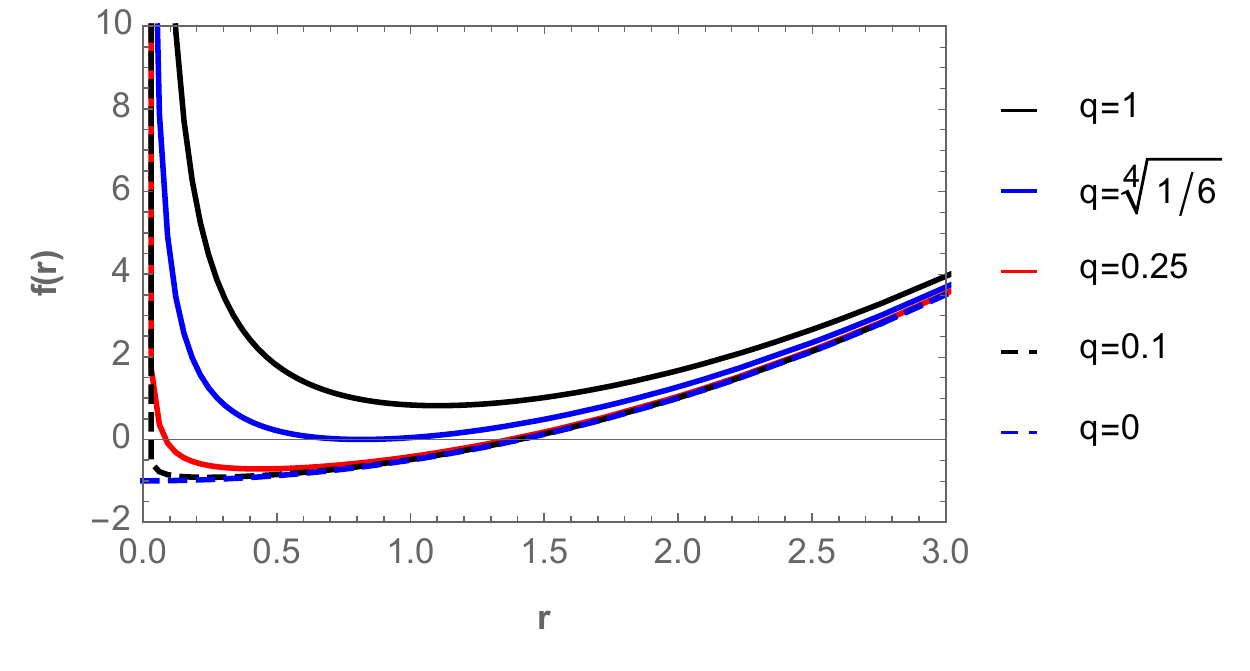}
\includegraphics[width=0.45\textwidth]{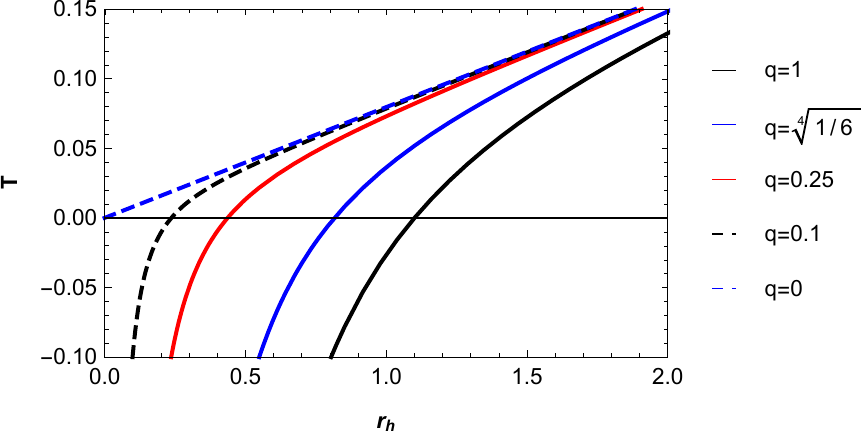}
\end{center}
\caption{Lapse function behavior as a function of $r$ (top panel), and temperature as function of the event horizon $r_h$ (bottom panel). Here, we consider $\Lambda=-0.5$ and different values of electric charge $q$. When $q=0$, the temperature profile of BTZ black hole is recovered, and $q=(1/6)^{1/4}$ corresponds to the extremal case.}
\label{function0}
\end{figure}

\begin{figure}[!h]
\begin{center}
%\plotone{Figg2}
\includegraphics[width=0.40\textwidth]{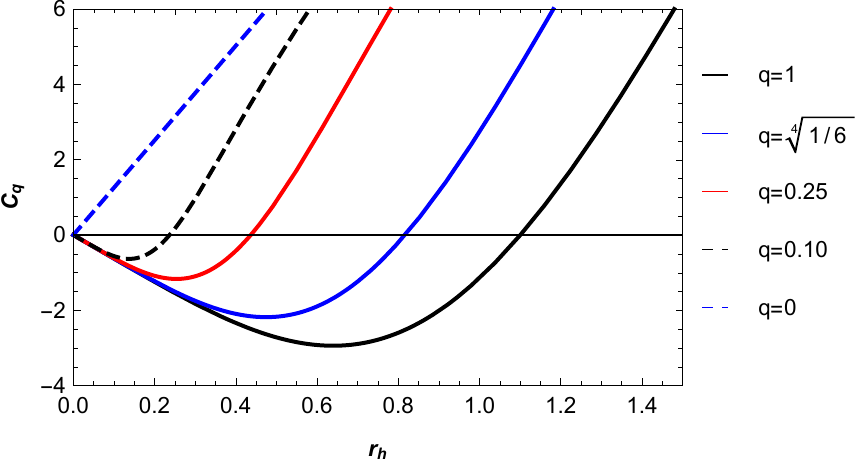}
\end{center}
\caption{Heat capacity as function of event horizon $r_h$. Here, we consider $\Lambda=-0.5$ and different values of electric charged $q$. When $q=0$, we recover the heat capacity profile of a static BTZ black hole.} 
\label{heat}
\end{figure}

We begin by determining the entropy of the geometry, which is assumed to satisfy the Bekenstein-Hawking entropy space, i.e, $S=\frac{A}{4}=4 \pi r_{h}$. {\b{There}} we can obtain the mass parameter as a function of entropy and charge. Ref. \cite{Kastor:2009wy} suggests that the mass $M$ of an AdS black hole can be interpreted as enthalpy in classical thermodynamics, rather than the total energy of the spacetime $M=H(S,q)$. This point will be important later when using the cosmological constant as variable. Meanwhile, from the mass of the black hole, Eq. (\ref{f(r)}) yields the following equation
\begin{equation}
 M(S,q)=- \frac{\Lambda S^2}{16 \pi^2}+\frac{16 \pi q^2}{3S}, \label{M}
\end{equation}
and using energy conservation, or the first law of black hole mechanics, we have
\begin{equation}
 dM=TdS+\Phi dq, \label{1law}
\end{equation}
and then we can obtain the thermodynamic variables as the temperature 
\begin{equation}
 T=\left(\frac{\partial M}{\partial S}\right)_{\Phi}=-\frac{\Lambda S}{8\pi^2}-\frac{16 \pi q^2}{3 S^2}\,, \label{T}
\end{equation}
and the electric potential
\begin{equation}
 \Phi= \left(\frac{\partial M}{\partial q}\right)_{T}=\frac{32 \pi q}{3S}\,. 
\end{equation}
From Eq. (\ref{T}), the temperature is positive when 
\begin{equation}
3S^3+\frac{128\pi^3q^2}{\Lambda}>0\,, \label{con}
\end{equation}
which is a standard requirement of black hole mechanics. At $q=0$, we obtain the well-known result $S_{BTZ}>0$. 
%Using the definition of the entropy, as was computed above, $S=4 \pi r_h$ and we see 
Note that the temperature is vanished for $r_h=r_{extrem}=\left(- \frac{2}{3} \frac{q^2}{\Lambda}\right)^{\frac{1}{3}}$, and for $r_h<r_{extrem}$ we obtain a negative temperature; therefore, this is a region with nonphysical meaning where the thermodynamics description breaks down, see Fig. \ref{function0}.
In order to understand this result and study the thermodynamic stability of this solution, we compute heat capacity from Eq. (\ref{M}), and move toward understanding of the possible critical behavior of this 2+1 AdS black hole
\begin{equation}
 C_q= T \left(\frac{\partial S}{\partial T}\right)_{q}=-S \left(\frac{128 \pi^3 q^2-3 \Lambda S^3 }{256 \pi^3 q^2+3 \Lambda S^3 }\right). \label{cq}
\end{equation}
Now, from Eqs. (\ref{T}) and (\ref{cq}), we can see that the temperature and the heat capacity are always positive when the condition (\ref{con}) is satisfied. These results imply that a $2+1$ Coulomb-like black hole with a positive definite temperature must be a stable thermodynamic configuration. Fig. \ref{heat} shows the heat capacity for different values of electric charge as a function of event horizon. Similar solutions and description for rotating and charged BTZ black holes can be found in Ref. \cite{Quevedo:2008xn} and \cite{Akbar:2011qw,Frassino:2015oca,Dehghani:2016agl}.
Now, according to \cite{Davies}, a change in the sign of heat capacity suggests an instability or a phase transition among the black hole configurations. We will study this point in more detail below; however, the main conclusion of Davies's approach establishes the correlation among drastic change in stability, properties of a thermodynamic black hole system, and a change in sign of heat capacity. In brief, a negative heat capacity represents a region of instability, whereas the stable domain is characterized by a positive heat capacity.
Indeed, it is well-described that the canonical ensemble of black holes resolve to a locally thermodynamic stable system if its heat capacity is positive or non-vanishing. Therefore, at points of vanishing or divergent heat capacity, there is a first or second-order phase transition, respectively. Therefore, we need to determine the positivity of heat capacity $C_q>0$ or also the positivity of $\partial{S}/\partial{T}$ or ($\partial
^2{M}/\partial{S}^2$) with $T>0$ as sufficient conditions to ensure the local stability of the black hole. 
\begin{figure}[!h]
\begin{center}
%\plotone{Figg2}
%\includegraphics[width=0.6\textwidth]{Figg3.eps}
\includegraphics[width=0.45\textwidth]{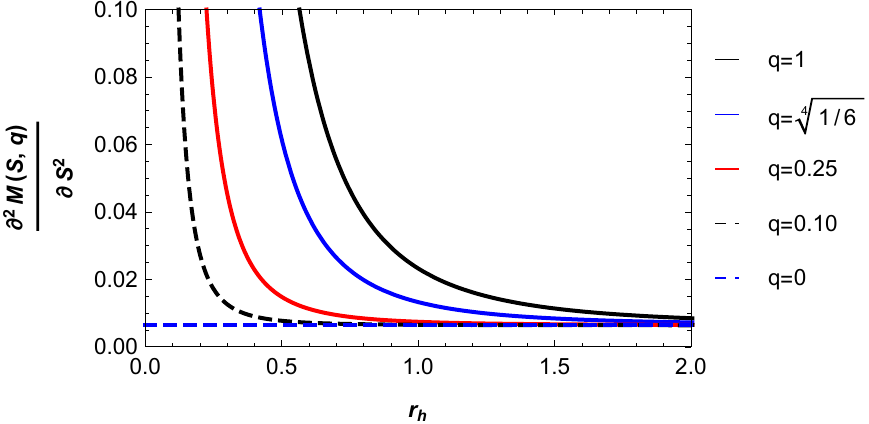}
\end{center}
\caption{This plot shows $\frac{\partial ^2 M(S,q)}{\partial S^2}$ as function of event horizon $r_h$. Here, we consider $\Lambda=-0.5$ and different values of electric charge $q$. When $q=0$, the $\frac{\partial ^2 M(S,q)}{\partial S^2}$ profile of a BTZ black hole is obtained.} \label{function1}
\end{figure}
 Figs. \ref{function0}, \ref{heat}, and \ref{function1} show that this black hole solution is locally stable from a thermal point of view - i.e., the heat capacity $C_q$ is positive and free of divergent terms. Therefore, the heat capacity is a regular function for all real positive values where $r>r_h$, and has positive $\frac{\partial ^2 M(S,q)}{\partial S^2}$. We will return to this point in more detail in the 
 next section.
 \subsection{Comparison with charged BTZ black holes}
\begin{figure}[!h]
\begin{center}
%\plotone{Figg2}
\includegraphics[width=0.4\textwidth]{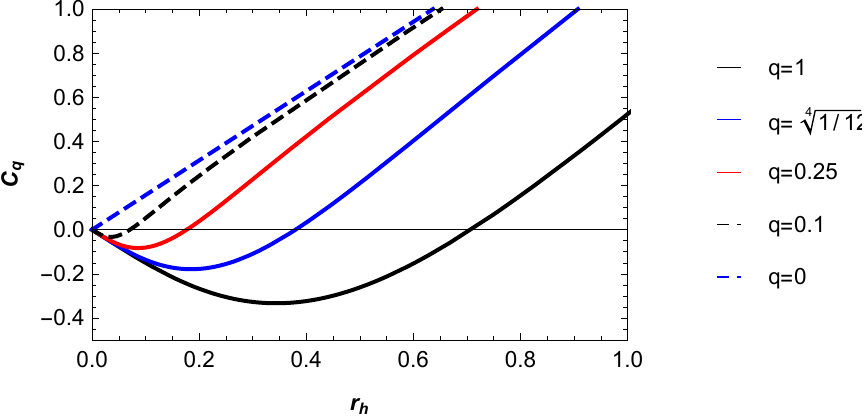}
\end{center}
\caption{Heat capacity as function of the event horizon $r_h$. Here, we consider $\Lambda=-0.5$ and different values of electric charge $q$. When $q=0$, we recover the heat capacity profile of static BTZ black holes.} 
\label{heatBTZ}
\end{figure}

 In this section, we establish the differences or similarities in thermodynamics descriptions of charged BTZ black holes - where $m$ is an integration constant related to the black hole mass $M$ through $M=\frac{m}{4}$ (\ref{BTZ}) - and Coulomb-like black holes with nonlinear electrodynamics interaction. We present the main results obtained in Refs. \cite{Frassino:2015oca,Dehghani:2016agl}. Starting from enthalpy $H(S,q)=M(S,q)$ of the charged BTZ black hole, which is given by
 \begin{equation}
 M(S,q)=\frac{1}{16}\left(\frac{8S^2|\Lambda|}{\pi^2}-q^2Log\left(\frac{2S \sqrt{|\Lambda|}}{\pi}\right)\right)\,, \label{MBTZ}
 \end{equation}
 the temperature then yields
\begin{equation}
 T=\left(\frac{\partial M}{\partial S}\right)_{\Phi}=\frac{\Lambda S}{\pi^2}-\frac{q^2}{16S}\,, \label{TBTZ}
\end{equation}
and is positive when 
\begin{equation}
16S^2 \Lambda-\pi^2q^2>0, \label{conBTZ}
\end{equation}
which is a standard requirement of black hole mechanics. For $q=0$, we obtain the well known result $S_{BTZ}>0$. Using the definition of entropy $S=\frac{\pi}{2} r_h$, we obtain vanished temperature for $r_h=r_{extrem}$; and, for $r_h<r_{extrem}$, a negative temperature. Therefore, there is a region with nonphysical meaning where the thermodynamics description breaks down.
\begin{figure}[!h]
\begin{center}
%\plotone{Figg1}
%\includegraphics[width=0.45\textwidth]{metricB.pdf}
\includegraphics[width=0.45\textwidth]{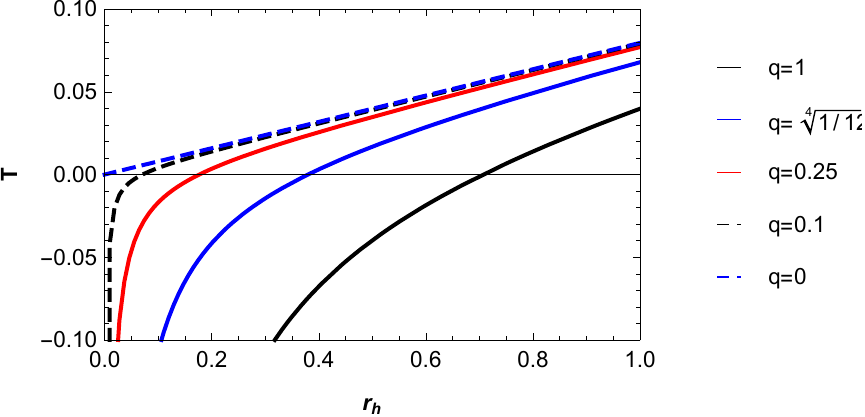}
\end{center}
\caption{Temperature as a function of event horizon $r_h$ (right panel) for a charged BTZ black hole. Here, we consider $\Lambda=-0.5$ and different values of electric charge $q$. When $q=0$, the temperature profile of BTZ black holes is recovered.}
\label{function3}
\end{figure}
As in the nonlinear Coulomb-like case, in order to understand this result and study the thermodynamics stability of this solution, we compute heat capacity from (\ref{MBTZ}) and move toward understanding possible critical behavior, followed by a comparison of our results with charged BTZ black holes.
\begin{equation}
 C_q= T \left(\frac{\partial S}{\partial T}\right)_{q}=-S \left(\frac{\pi^2 q^2-16 \Lambda S^2 }{\pi^2 q^2+16 \Lambda S^2}\right). \label{cqBTZ}
\end{equation}
Now, from Eqs. (\ref{TBTZ}) and (\ref{cqBTZ}), the temperature and the heat capacity are always positive when condition (\ref{conBTZ}) is satisfied. This implies that the $2+1$ charged BTZ black hole with a positive definite temperature must be a stable thermodynamic configuration. Fig. \ref{heatBTZ} shows the heat capacity for different values of electric charge as a function of the event horizon. Similar solutions and descriptions for rotating and charged BTZ black holes can be found in Ref. \cite{Quevedo:2008xn} and \cite{Akbar:2011qw,Frassino:2015oca,Dehghani:2016agl}.
Thus we confirm that there are no major differences in the standard thermodynamics descriptions between charged BTZ black holes and Coulomb-like $2+1$ dimensional black holes: both have thermodynamic stability. We have shown in this section that a black hole under nonlinear electrodynamics in $2+1$ dimensions with a Coulomb-like potential is a stable thermodynamic configuration, at least using canonical ensemble descriptions (for a discussion of ensemble dependency in charged BTZ black holes, see \cite{Hendi:2015wxa} cite{Hendi:2010px}). 

\subsection{The geometrothermodynamics approach}
\begin{figure}[!h]
\begin{center}
%\plotone{Figg2}
\includegraphics[width=0.7\textwidth]{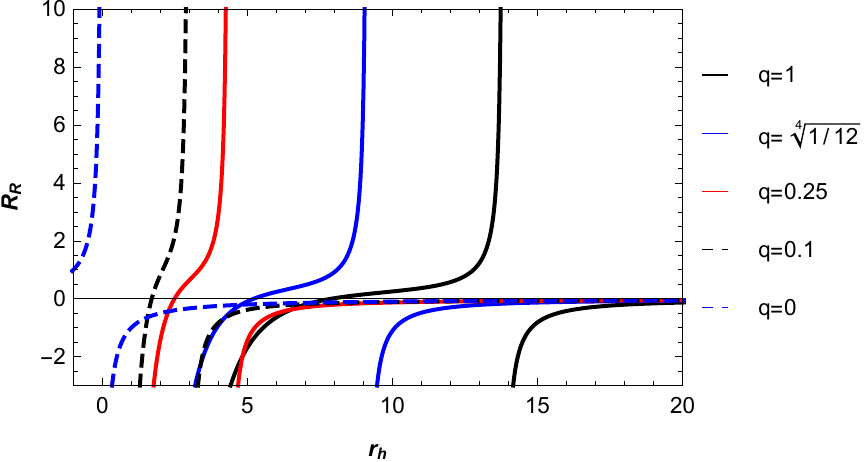}
\end{center}
\caption{$R_R$ as a function of the event horizon $r_h$. Here, we consider $\Lambda=-0.5$ and different values of electric charge $q$. When $q=0$, the $R_R$ profile of a BTZ static black hole is obtained.} \label{R}
\end{figure}

This section describes the essential aspects of geometry in thermodynamic phase space. The geometrical approach - refined in Ruppeiner \cite{Ruppeiner:1995zz} and Weinhold \cite{W} - has been extensively applied to black hole thermodynamics, e.g., \cite{Ferrara:1997tw,Aman:2005xk,Sarkar:2006tg,Shen:2005nu,Quevedo:2008xn,Banerjee:2010bx,Astefanesei:2010bm,Wei:2010yw,Liu:2010sz,Niu:2011tb,Wei:2012ui,Banerjee:2016nse,Vetsov:2018dte,Dimov:2019fxp,Bhattacharya:2019awq, Han:2011zzd, Han:2012hc}. This approach builds an analogous space of thermodynamics parameters, and then defines appropriate metric tensors for this space for which line elements measure the distance between two neighbouring fluctuation states in the state space. 
Other geometric quantities, such as curvature, represent thermodynamics properties and critical behavior in black hole systems. Particularly, this thermodynamic curvature provides information on the nature of the interaction among the fundamental properties constituent of the system. It is well known, for example, that the thermodynamics parameter space for an ideal gas is flat with vanished curvature, reflecting the nature of a collection of non interacting particles. For Van der Waals fluids, in contrast, we find a curved thermodynamics space, implying an attractive or repulsive interacting system for positive or negative curvature, respectively. Importantly, geometrothermodynamics allows for singularity structures, whose presence indicates phase transitions in the system. 
Let us start with the Weinhold metric, defined by the second derivatives of 
internal energy with respect to entropy and other extensive parameters $(S,q)$
\begin{equation}
 g^W_{bc}=\frac{\partial ^2 M(X^a)}{\partial X^b \partial X^c }\,,
\end{equation}
where $X^a=(S, x^{\hat {a}})$; $S$ represents entropy; and $x^{\hat {a}}$, all other extensive variables. Next, taking the Ruppeiner metric - defined as the second derivatives of entropy with respect to internal energy and other extensive parameters - we have 
\begin{equation}
 g^R_{bc}=\frac{\partial ^2 S(Y^a)}{\partial Y^b \partial Y^c }\,,
\end{equation}
where $Y^a=(M, y^{\hat a} )$; $M$ represents mass; and $y^{\hat a}$, all other extensive variables. These two metrics are conformally related by $ds^2_R=\frac{1}{T} ds^2_W$.
We may compute both metrics in their natural coordinates. The Weinhold line element is given by
\begin{widetext}
\begin{equation}
 ds^2_W=\left(\frac{256\pi^3 q^2-3\Lambda S^3}{24 S^3 \pi^2}\right)dS^2-\frac{64 \pi q}{3S^2}dSdq+\frac{32 \pi}{3S}dq^2\,.
\end{equation}
\end{widetext}
 Similar to the condition of positivity for $\frac{\partial ^2 M(S,q)}{\partial S^2}$ in black hole thermodynamic stability, so too are there conditions for the Weinhold metric, essentially a Hessian matrix constructed with the following mass formula (\ref{M})
\begin{equation}
 \mathbf{H}^M_{S,q}=\left(\begin{matrix}
\frac{256\pi^3 q^2-3\Lambda S^3}{24 S^3 \pi^2} & -\frac{64 \pi q}{3S^2}\\ \\
-\frac{64 \pi q}{3S^2} & \frac{32 \pi}{3S}
\end{matrix}\right), \label{matriz}
\end{equation}
 from which we can determine thermodynamic stability of the grand canonical ensemble using standard extensive parameters entropy and charge, using determinant $det( \mathbf{H}^M_{S,q})=-\frac{4 \left(256 \pi ^3 q^2+\Lambda S^3\right)}{3 \pi S^4}$. The grand canonical ensemble of this black hole is always stable; however, we know from Eq. (\ref{cq}) that there is at least one region of instability for $C_q=0$, and another for $C_q<0$; and from Davies's approach that there is a first-order phase transition. This is evidence of ensemble dependency. Next, the Weinhold thermodynamics curvature scalar is given by
\begin{equation}
 R_W=-\frac{4 \pi ^2}{\Lambda S^2}\,,
\end{equation}
which is regular and positive.
 Let us explore different ways to fix this apparent conflict, first, by computing the Ruppeiner curvature; and second, by considering the extended thermodynamics space 
% and consider 
where the cosmological constant is another thermodynamics variable. The Ruppeiner line element is given by 
 \begin{widetext}
\begin{equation}
 ds^2_R=-\frac{1}{S}\left(\frac{256 \pi^3 q^2-3S^3 \Lambda}{128 \pi^3 q^2+3S^3 \Lambda}\right)dS^2+\left(\frac{512 \pi^3 q }{128 \pi^3 q^2+3S^3 \Lambda}\right)dSdq-
 \left(\frac{256 \pi^3 S }{128 \pi^3 q^2+3S^3 \Lambda}\right)dq^2\,.
\end{equation}
\end{widetext}
Thermodynamic stability occurs when metric fluctuation is positive, written as $g^R_{SS}>0$, $g^R_{qq}>0$ and $det(g_R)>0$. The first two conditions are satisfied at all points of $(S,q)$ space, except for where the heat capacity becomes vanishing; and the last condition, at $det(g_R)=-\frac{768 \left(256 \pi ^6 q^2+\pi ^3 \Lambda S^3\right)}{\left(128 \pi ^3 q^2-3 \Lambda S^3\right)^2}$. To determine the existence of a phase transition in the system at this point, we consider the Ruppeiner thermodynamics curvature scalar
\begin{equation}
 R_R=\frac{64 \pi ^3 q^2}{3 \Lambda S^4}+\frac{384 \pi ^3 q^2}{128 \pi ^3 q^2 S-3 \Lambda S^4}-\frac{1}{S}\,,
\end{equation}
that presents a genuine divergence at %the point where ratio
\begin{equation}
128 \pi ^3 q^2-3 \Lambda S^3=0\,,
\end{equation}
i.e., when $r_h=r_{extrem}$ 
%it 
is satisfied, see Fig (\ref{R}). This is precisely the point where heat capacity becomes vanishing. This represents microscopic interacting behavior, confirms a phase transition in this black hole solution, and shows that the Ruppeiner thermodynamic curvature correctly describes the transition from a region with positive and well-defined temperature to a region with a nonphysical negative temperature.
%In Fig (\ref{R}) we plot the behavior of the Ruppeiner scalar as a function of the event horizon.

In sum, we can see that the Coulomb-like black hole solution considering nonlinear Born-Infeld electrodynamics in $2+1$ dimensions shows two horizons, with a phase transition to the extremal solution at the point $r_{extrem}=\left(- \frac{2}{3} \frac{q^2}{\Lambda}\right)^{\frac{1}{3}}$ where, for distance less than $r_{extrem}$ - the vanishing temperature limit, or the geometrical extremal limit - the heat capacity is vanishing and negative, and temperature is therefore nonphysical. In this sense, the unstable region ($C_q < 0$) corresponds to a nonphysical region with negative temperature. According to Davies, a heat capacity sign change indicates a strong change in thermodynamics system stability, while negative heat capacity represents a region of instability with negative temperature. Because the third law of thermodynamics imposes a positive temperature, the description breaks down at the extremal limit. Similar results were obtained for
 the BTZ black hole \cite{Quevedo:2008xn}, \cite{Akbar:2011qw}. Additionally, these results broadly imply for geometrothermodynamics that, while the Weinhold metric is unable to demonstrate system divergence and therefore phase transition, the Ruppeiner metric correctly showed true curvature divergence or singularity at $r_{extrem}=\left(- \frac{2}{3} \frac{q^2}{\Lambda}\right)^{\frac{1}{3}}$. The capability of one metric over another for Kerr black holes was similarly found in Ref. \cite{Quevedo:2007ws}.\\

 \subsection{Comparison with charged BTZ black holes}
In this case, the Ruppeiner element is given by 
\begin{widetext}
\begin{equation}
 ds^{2 BTZ}_{R}=-\frac{1}{S}\left(\frac{ \pi^2 q^2+16S^2 \Lambda}{\pi^2 q^2-16S^2 \Lambda}\right)dS^2+\left(\frac{2 \pi^2 q }{ \pi^2 q^2-16S^2 \Lambda}\right)dSdq-
 \left(\frac{2 \pi^2 S Log\left(\frac{2 S \sqrt{|\Lambda|}}{\pi}\right) }{\pi^2 q^2-16S^2 \Lambda}\right)dq^2.
\end{equation}
\end{widetext}
Again, positive metric fluctuation implies stability, which is so for $g^R_{SS}>0$, $g^R_{qq}>0$ and $det(g_R)>0$. Again are the first two conditions satisfied for all points of $(S,q)$ space, except where heat capacity becomes vanishing; and the last condition, by
\begin{widetext}
$$
det(g_R^{BTZ})= -\frac{2 \pi ^2 \left(\pi ^2 q^2 \log \left(\frac{2 \sqrt{\Lambda } S}{\pi }\right)+2 \pi ^2 q^2+16 \Lambda S^2 \log \left(\frac{2 \sqrt{\Lambda } S}{\pi }\right)\right)}{\left(\pi ^2 q^2-16 \Lambda S^2\right)^2},
$$
\end{widetext}
Similarly as above, this may imply a phase transition in the system. The 
Ruppeiner thermodynamics curvature \cite{Xu:2020ftx} is given by,
\begin{widetext}
\begin{equation}
 R_R^{BTZ} =\frac{A+B * \log \left(\frac{2 \sqrt{\Lambda } S}{\pi }\right)+C* \log ^2\left(\frac{2 \sqrt{\Lambda } S}{\pi }\right)}{\pi S \left(\pi ^2 q^2-16 \Lambda S^2\right) \left(2 \pi ^2 q^2 \log \left(\frac{2 \sqrt{\Lambda } S}{\pi }\right)+q^2+32 \Lambda S^2 \log \left(\frac{2 \sqrt{\Lambda } S}{\pi }\right)\right)^2},
\end{equation}
\end{widetext}
where constants $A$, $B$, $C$ are 
\begin{widetext}
\begin{eqnarray*}
A&=&256 \pi ^3 \Lambda ^2 q^2 S^4+1024 \pi ^2 \Lambda ^2 q^2 S^4-1024 \pi \Lambda ^2 q^2 S^4+\\&+&16 \pi ^5 \Lambda q^4 S^2-64 \pi ^4 \Lambda q^4 S^2-256 \pi ^3 \Lambda q^4 S^2+512 \pi ^2 \Lambda q^4 S^2-\pi ^7 q^6+4 \pi ^5 q^6-4096 \pi \Lambda ^3 S^6,\\
B&=&-256 \pi ^3 \Lambda ^2 q^2 S^4+3072 \pi ^2 \Lambda ^2 q^2 S^4-144 \pi ^5 \Lambda q^4 S^2+\\&+&224 \pi ^4 \Lambda q^4 S^2+256 \pi ^3 \Lambda q^4 S^2+\pi ^7 q^6-28672 \pi \Lambda ^3 S^6+24576 \Lambda ^3 S^6,\\
C&=&2048 \pi ^3 \Lambda ^2 q^2 S^4+96 \pi ^5 \Lambda q^4 S^2+8192 \pi \Lambda ^3 S^6.
\end{eqnarray*}
\end{widetext}
Note that the Ruppeiner scalar shows a true curvature divergence or singularity at $r_h=r_{extrem}$. 
%similar situation was reported in Ref. \cite{Xu:2020ftx}.}} 
In the next section,
%, we would like to mention our goal for next section, because it is possible
we will consider 
%to consider 
the cosmological constant
%as we mention above, 
as an additional extensive thermodynamic variable; indeed,
%with a the thermodynamics equilibrium three-dimensional space, the in the literature
%that is called 
the cosmological constant is treated as a pressure term in extended thermodynamics   \cite{Kastor:2009wy,Dolan:2010ha,Dolan:2011xt,Gunasekaran:2012dq}.

\section{Extended thermodynamics description}
\label{ETD}

The heat capacity expression Eq.(\ref{cq}) and the Hessian matrix determinant Eq. (\ref{matriz}) are incompatible with the change of phase transition proposed by Davies, and so we have ensemble dependence. In resolving this, we may extend the thermodynamics space and consider the cosmological constant as another thermodynamics variable. Here, the standard extensive parameters will be entropy, charge, and the cosmological constant. We consider the cosmological constant a source of dynamic pressure using the relation 
$P=-\frac{\Lambda}{8 \pi}$ \cite{Dolan:2010ha, Dolan:2011xt}. Now, understanding black hole mass as the enthalpy, all thermodynamic descriptions are given by functions $H=H(S,P,q)$ as follows
\begin{equation}
 H=M(S,P,q)=\frac{PS^2}{2\pi}+\frac{16 \pi q^2}{3 S}\,.
\end{equation}
The pressure-related conjugate quantity is thermodynamic volume,
\begin{equation}
 V=\left(\frac{\partial H}{\partial P}\right)_{S,q}=8 \pi r^2_h\,.
\end{equation}
The first law of black hole thermodynamics in the extended phase space reads as 
\begin{equation}
 dH=T dS+PdV+\Phi dq\,,
\end{equation}
 
Before continuing, it is worthwhile to analyze a (2 + 1)-dimensional BTZ black hole under extended thermodynamics to establish important concepts and features. Previously studied in Ref. \cite{Dolan:2011xt}, the main results give black hole mass by
\begin{equation}
 M=H(S,P)=\frac{4PS^2}{\pi}\,, \label{BTZstatic}
\end{equation}
where entropy was defined as $S=\frac{A}{4}$ and black hole area $A=2 \pi r_h$. The following equation of state was obtained
\begin{equation}
 P\sqrt{V}=\frac{\sqrt{\pi}T}{4}\,, \label{BTZstaticEoS}
\end{equation}
which corresponds to that of an ideal gas. That author concluded that static BTZ black holes are associated with noninteracting microstructures. Next, for rotating BTZ black hole, as presented in Refs. \cite{Ghosh:2020kba}, \cite{Frassino:2015oca} \cite{Gunasekaran:2012dq}, the equation of state is given by,
\begin{equation}
 P=\frac{T}{v}+\frac{8J^2}{\pi v^4}\,, \label{RBTZEoS1}
\end{equation}
where $v=4 r_h$ is the specific volume. This can be interpreted as a Van der Waals fluid, as given by \cite{Kubiznak:2012wp}
\begin{equation}
 \left(P+\frac{a}{v^2}\right)(v-b)=kT\,, \label{VdW}
\end{equation}
where $v$ is the specific volume and $k$ is the Boltzmann constant. Eq. (\ref{VdW}) describes an interacting fluid with critical behavior, and so rotating BTZ black holes are repulsive and do not have any critical thermodynamics behavior. Next, in studying lower dimensional black hole chemistry for charged and rotating BTZ black holes, Ref. \cite{Frassino:2015oca} tested the reverse isoperimetric inequality \cite{RII} under the conjecture that its ratio
\begin{equation}
 \Re=\left(\frac{(D-1)V}{\omega_{D-2}}\right)^{\frac{1}{D-1}}\left(\frac{\omega_{D-2}}{A}\right)^{\frac{1}{D-2}}\,, \label{RII}
\end{equation}
always satisfies $\Re \geq 1$ for conjugate thermodynamics volume $V$, and the horizon area $A$, where $\omega_d=\frac{2 \pi^{\frac{d+1}{2}}}{\Gamma(\frac{d+1}{2})}$ corresponds to the area of a d-dimensional unit sphere. Those authors found that the rotating case for $\Re=1$ results in a saturated reverse isoperimetric inequality, i.e., that rotating BTZ black holes have maximal entropy. 
Ref. \cite{Frassino:2015oca,Gunasekaran:2012dq} give, for the charged BTZ black hole, the mass as 
\begin{equation}
 M=H(S,P)=\frac{4PS^2}{\pi}-\frac{q^2}{32}log\left(\frac{32PS^2}{\pi}\right)\,, \label{BTZChargedd}
\end{equation}
where entropy is $S=\frac{\pi}{2}r$; and the equation of state as \cite{Gunasekaran:2012dq}
\begin{equation}
 P=\frac{T}{v}+\frac{q^2}{2\pi v^4}\,, \label{RBTZEoS}
\end{equation}
where $v=4 r_h$ is the specific volume. Therefore, the charged BTZ black hole does not show critical behavior. Ref. \cite{Frassino:2015oca} also discussed the violation of the reverse isoperimetric inequality for $\Re<1$, for which charged BTZ black holes are always superentropic. For similar discussions on different cases of AdS black holes, see \cite{Zou:2013owa, Kubiznak:2012wp,Altamirano:2014tva,Cong:2019bud,Zhou:2020vzf,Chaturvedi:2017vgq,Mir:2019ecg,Belhaj:2015hha}.

 Now, returning to the $2+1$ nonlinear Coulomb-like black black hole, temperature and the respective equation of states is as follows 
\begin{equation}
 P=\frac{\sqrt{\pi} T}{\sqrt{2 V}}+\frac{4 \sqrt{2\pi} q^2}{3 V^{3/2}}\,,
\end{equation}
which may be interpreted as a Van der Waals fluid. Therefore, the $2+1$ Coulomb-like black hole is associated with repulsive microstructures, consistent with non-vanishing Ruppeiner curvature and non-critical thermodynamics behavior. As such, the black hole does not have phase transitions. At $q=0$, we obtain an equation of state similar to a static BTZ black hole, $P \sqrt{V}\propto T$. In computing the 
%Reverse 
reverse isoperimetric inequality, we obtain $\Re = \sqrt{2 \pi}>1$. As mentioned previously, the nonlinear electrodynamics interaction does not require additional regulating terms.

\begin{itemize}
\item Holographic heat engine
\end{itemize}

\begin{figure}[!h]
\begin{center}
%\plotone{Figg2}
\includegraphics[width=0.7 \textwidth]{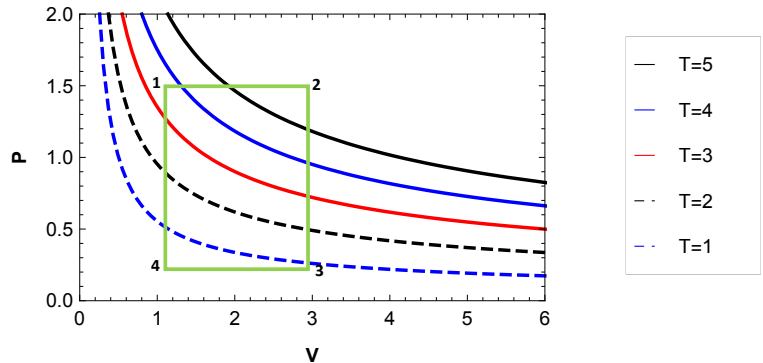}
\end{center}
\caption{Isothermal curves for charged Nonlinear Coulomb-like black holes.}
\label{function000}
\end{figure}

Our analysis of efficiency in $2+1$ nonlinear black holes as holographic heat engines follows that of PV criticality as in Refs. \cite{Hennigar:2017apu,Johnson:2014yja,Johnson:2016pfa,Johnson:2015fva,Guo:2019rdk,Yerra:2018mni}. Ref. \cite{Johnson:2014yja} defined holographic heat engines via an analogous extraction of mechanical work from heat energy. Taking the extended First Law of black hole thermodynamics, which includes the $PdV$ term in $ dH=T dS+PdV+\Phi dq$, the working substance is a black hole solution of the gravity system with volume, pressure, temperature, and entropy. To begin, take the equation of state (function of $P(V,T)$) where the
engine is a closed path in the P-V plane of net input heat flow $Q_H$ and net output heat flow $Q_C$, such that $Q_H = W + Q_C$. It is well-known in classical thermodynamics that heat engine efficiency is $\eta=\frac{W}{Q_H}=1-\frac{Q_H}{Q_C}$. Some classic cycles call for isothermal expansion and compression at temperatures $T_H$ and $T_C$ ( $T_H > T_C$ ).
We can show net heat flows along each isobar by
\begin{equation}
 Q=\int _{T_i}^{T_f} C_P dT\,,
\end{equation}
 and so mechanical work is computed by $W= \int PdV$. In classical thermodynamics, the Carnot cycle - which takes two pairs of isothermal and adiabatic processes - has the highest efficiency, given by 
 $\eta=1- \frac{T_C}{T_H}$. Following the construction of a black hole heat engine in Ref. \cite{Johnson:2016pfa}, we define a simple heat cycle with isotherm pairs at high $T_H=T_1$ and low $T_C=T_2$ temperatures, connected through isochoric paths. As in isothermal expansion and compression in the Carnot cycle, heat absorbed is $Q_H$, and discharged, $Q_C$ (Fig. \ref{function000}). The efficiency of this cycle is given by the simple expression 
\begin{equation}
 \eta=1-\frac{M_3-M_4}{M_2-M1}\,,
\end{equation}
In cycling along isochoric paths $V_1=V_4$ and $V_2=V_3$ and isobaric paths $P_1=P_2$ and $P_3=P_4$, engine efficiency is given by
\begin{equation}
 \eta=3\frac{(P_1-P_4)S_1S_2(S_2+S_1))}{3P_1(S_2+S_1)S_1S_2-32\pi q^2}, \label{eta}
\end{equation}
A similar result for nonlinear electrodynamics black holes was found in Ref. \cite{Balart:2019uok}. Note that our expression for the limit $q=0$ is
%, we obtain the efficient of BTZ static black hole at the leading order as 
\begin{equation}
 \eta=1- \frac{T_C}{T_H} \sqrt{\frac{V_2}{V_4}}\,,
\end{equation}
and is also consistent with the result reported in Ref. \cite{Mo:2017nhw} for static BTZ black holes.
%\newpage
\section{Concluding comments}
\label{conclusion}
In this paper we have studied the thermodynamics description of $2+1$ dimensional Coulomb-like black holes under nonlinear electrodynamics and with a traceless energy-momentum tensor in $2+1$ dimensions. Remarkably, this solution was obtained for a circularly symmetric static metric describing an asymptotically anti-de Sitter black hole in a Coulomb-like field.
Notably - and in contrast with charged BTZ solutions, which diverge and yield quasilocal mass due to its logarithmic term - our derived charged black holes were shown to possess finite mass. In short, our method of thermodynamics analysis does not require renormalization.
Our solution shows stable thermodynamic behavior: in all regions where $r_h>r_{extrem}$, temperature and heat capacity $C_q$ are always positive and free of singular points, indicative of a stable black hole thermodynamics configuration where no phase transitions occur. Although Davies's approach would suggest the presence of a first-order phase transition for region $r_h<r_{extrem}$, where $C_q<0$, the temperature in this region is negative - and nonphysical - and so the thermodynamics description breaks down. In contrasting these results, we used thermodynamics phase space geometry curvatures via Weinhold and Ruppeiner metrics. Of these, the Weinhold metrics at the grand canonical ensemble conclude a stable, divergence-free black hole, similar to the heat capacity canonical ensemble. Second, we obtained a non-vanishing Rupeinner's curvature, indicating 
an interacting system; however, a divergent point where heat capacity becomes vanishing was found. Thus, we confirmed that this black hole solution has a first-order phase transition, but that the region where $C_q<0$ is nonphysical. In this sense, the unstable region corresponds to a nonphysical region with negative temperature. According to Davies, a heat capacity sign change indicates a strong change in thermodynamics system stability, while negative heat capacity represents a region of instability with negative temperature. Because the third law of thermodynamics imposes a positive temperature, the description breaks down at the extremal limit.
Finally, regarding implications of this thermodynamic stability, we considered the cosmological constant as source of a dynamical pressure using the relation $P = - \frac{\Lambda}{8 \pi}$ using the enthalpy function $H = H(S, P, q)$. Employing the first law of black hole mechanics, we computed the equation of state 
\begin{equation}
 P=\frac{\sqrt{\pi} T}{\sqrt{2 V}}+\frac{4 \sqrt{2\pi} q^2}{3 V^{3/2}}\,, \label{qq}
\end{equation}
 which can be interpreted as a Van der Waals fluid. Therefore we concluded that the $2+1$ dimensional Coulomb-like black hole is associated with repulsive microstructures, consistent with non-vanishing Ruppeiner curvature. From Eq. (\ref{qq}) and its graphic in Fig. \ref{R}, we concluded that there is no critical thermodynamics behavior, and that there are no phase transitions. Here, for $q=0$, we obtain an equation of state similar to that of a static BTZ black hole, $P \sqrt{V}\propto T$. The ratio of reverse isoperimetric inequality was calculated as $\Re = \sqrt{2 \pi}>1$. As mentioned previously, the nonlinear electrodynamics interaction does not require additional regulating terms. Finally, we constructed a simple heat cycle engine in the background of this black hole - with isotherm pairs at high $T_H = T_1$ and low $T_C = T_2$ temperatures connected through isochoric paths - similar to the Carnot cycle. The heat absorbed ($Q_H$) and discharged ($Q_C$) during isothermal expansion and compression gives a heat engine efficiency expression at limit $q=0$, or that of a static BTZ black hole.

\acknowledgments

%We would like to thank....
This work is supported by ANID Chile through FONDECYT Grant No 1170279 (J. S.). Y.V. would like to acknowledge support from the Direcci\'on de Investigaci\'on y Desarrollo at the Universidad de La Serena, Grant No. PR18142. B.W. was supported in part by NNSFC under grant No. 12075202.

\end{document}